\documentclass[showpacs,prl,twocolumn,groupedaddress,aps]{revtex4}
\usepackage{graphicx, epsfig}

\newcommand{\be}{\begin{equation}}
\newcommand{\ee}{\end{equation}}
\newcommand{\bea}{\begin{eqnarray}}
\newcommand{\eea}{\end{eqnarray}}

\begin{document}

\title{Is Eternal Inflation Eternal ?}

\author{ L.~Mersini-Houghton}
\email[]{mersini@physics.unc.edu}
\affiliation{Department of Physics and Astrononmy, UNC-Chapel Hill, NC, 27599-3255, USA\\ and,\\CITA, University of Toronto, Canada}

\date{\today}
 
\begin{abstract} 

In this paper we explore the relationship between the existence of eternal inflation and the initial conditions leading to inflation. We demonstrate that past and future completion of inflation is related, in that past-incomplete inflation can not be future eternal. Bubble universes nucleating close to the initial conditions hypersurface have the largest Lorentz boosts and experience the highest anisotropy. Consequently, their probability to collide upon formation is one. Thus instead of continuing eternally inflation ends soon after it starts. The difficulty in actualizing eternal inflation originates from the breaking of two underlying symmetries: Lorentz invariance and the general covariance of the theory which lead to an inconsistency of Einstein equations. Eternal inflation may not be eternal.
\end{abstract}


\maketitle

\subsection{1. Introduction}
Three current theories predict a multiverse extension of the Big Bang Inflationary Cosmology. They are: {\it i)} the many-worlds interpretation of quantum mechanics in conjunction with the decoherence mechanism; {\it ii)} the survival of high energy universes selected dynamically from the landscape of string theory; {\it iii)} eternal inflation whereby bubble universes continuously nucleate from the inflating background and collide with each other.

The existence of the multiverse would revolutionize physics. For this reason, investigating and testing theories of the multiverse is of fundamental importance. The first two theories above are related since the proposal to place the wavefunction of the universe on the landscape \cite{landscape, douglas} while addressing the wavefunction's decoherence \cite{laura}, embedds the many worlds interpretation of quantum mechanics into the string theory landscape via quantum cosmology. Observational tests of theory (${\it ii}$) were derived in \cite{tomo}, and three of the predictions made there have since been tested \cite{void,sigma8,darkflow}. Studies of observational signatures for theory (${\it iii}$) were conducted in \cite{aguirre}. In contrast to the theory of the universe from the landscape multiverse {\it ii} in which surviving universes have decohered and interact only via quantum entanglement, scenarios of eternal inflation {\it iii} expect their observational signatures to arise via collisions of bubble universes. (Some important subtleties of the issues that arise in relating theories (${\it ii}$) and (${\it iii}$) are clarified in the appendix).

In this first paper we investigate the nature of eternal inflation, specifically the consequences that initial conditions have on the continuation of inflation to future infinity. The effects of bubble collisions and the issue of instabilities related to fluctuations in bubble collisions will be presented elsewhere.

It is widely accepted that once inflation starts it continues generically to future infinity \cite{eternal}. Models of eternal inflation usually describe bubbles of lower energy ('true vacuum') that nucleate from the inflating background of a higher energy (false vacuum) $H^{2}_{F}$ at a nucleation rate per unit volume and unit time, given by $\lambda$. The interior space of one of these bubbles is assumed to describe our universe. The interior of the bubble thermalizes and grows with time. But, since the nucleation rate is small $\lambda H^{-4}_{F} \ll 1$ and the background energy is higher, then the volume of the inflating background grows faster than the volume covered by bubbles. There always exist inflating regions of 'false vacuum' from which new bubbles of 'true vacuum' can potentially nucleate, implying that spacetime consists of thermalized and inflating regions {\it ad infinitum}. Accordingly, although inflation switches on at some definite moment $t=t_i$ in the far past, it is expected to become future-eternal. The phase transition to a purely thermalized 'true vacuum' spacetime is never complete.  

Inflationary spacetimes are not past-complete \cite{BordeG, BordeV, Borde} in the sense that when inflation is traced back in time it can not be extended eternally to the past, in fact it can not extend beyond the surface of the initial conditions given by the initial time-slice $t=t_i$. The surface of the initial conditions is defined as the hypersurface at the moment $t=t_i$ when inflation first begins. Since the inflationary spacetime can not be continued beyond this singularity \cite{BordeV}, not only do initial conditions need to be specified ${\it apriori}$ but the time-slice $t = t_{i}$ provides a 'hard boundary' condition that cuts off the analytic continuation of the inflating spacetime region from the rest of the global DeSitter spacetime. With these assumptions, inflation has a beginning at the initial singularity but no end \cite{BordeG, BordeV, Borde}. Each bubble universe can reach the initial conditions boundary within a finite proper time \cite{BordeG}. 

One of the goals of eternal inflation is to make the haunting issue of its initial conditions, irrelevant for the interior cosmology of our bubble universe. It is thought that our bubble universe may not be sensitive to the choice of initial conditions for inflation if that moment of beginning is pushed far back in the past. But as we demonstrate here, past and future completion of inflation are related, i.e. eternal inflation scenarios that are past incomplete can not be future-eternal. In this light, the issue of initial conditions remains one of the most relevant topics not only for the interior cosmology of bubble universes but also for the existence of eternal inflation itself. If we think of two arbitrarily chosen events to the past and future of each other in the inflating background then the difference between them should be the redshifting by the scale factor $1/a(t)$. Apart from this scaling, the evolution equations that govern these events should have the same symmetries since the inflating background is a maximally symmetric space. Hence the selection of a 'preferred' time-slice, that of the initial conditions at $t =t_i$, breaks this Lorentz invariance symmetry and defines a preferred frame for the background, the gradient of the metric. But the congruence of geodesics determined by the flat metric of the inflationary phase converges at the initial conditions singularity. The symmetry breaking imprints itself on the bubble as an anisotropy of the initial conditions. The location dependent 'memory' of the initial conditions and the anisotropic distribution of bubbles \cite{GGV, Bucher} is a function of the boost factor of the observer with respect to the initial conditions surface.

Besides the anisotropy problem studied in \cite{GGV, Bucher}, bubbles that nucleate near the onset of inflation, i.e with a proper time $\tau \simeq 0$ from the surface, are subjected to a more serious problem. We show in Sec.3 they have a probability 1 to get hit and destroyed immediately upon formation, due to the convergence of geodesics near the initial conditions surface. Thus, the choice of the initial conditions for eternal inflation leads to a fast transition from inflating to thermalized spacetime, i.e. to the end of inflation. This effect is due to a maximum blueshifting of the velocities of observers near the initial conditions singularity, a scaling proportional to the convergence of the geodesics there. As will be shown in Sec.3, due to the observer's relativistic factor (rapidity) $\gamma$ diverging near the initial conditions, eternal inflation scenarios suffer an analogous transplanckian problem as the Hawking radiation with its blueshifted wavepackets near the horizon of a black hole.

It is well known that spacetimes with a preferred frame \cite{ted} can suffer pathological instabilities, one of which, the instability of eternal inflation, is studied here. We review the basics of bubble universes and eternal inflation in Sec.2, discuss the problem that initial conditions present for eternal inflation in Sec.3, and conclude in Sec.4.

\subsection{2. Brief Review: Eternal Inflation and Bubble Universes}

Single field open inflation is often described as a phase transition from false to true vacuum via bubble nucleation. The interior of a bubble is an open FRW universe. Since the universe is in a supercooled state and empty, models of eternal inflation assume that the interior of the bubble undergoes a second stage of inflation. The second stage of inflation in bubble's interior is usually realized by setting up the field to slow-roll down a potential
slope. The second stage of inflation inside the bubble is neccessary for diluting the curvature, monopoles and other dangerous relics as well as providing the mechanism for seeding the Cosmic Microwave Background (CMB) and
structure we observe in our universe. Such models are described in \cite{Linde, BucherT} for example.

Independently of the cosmology and the second stage of inflation in the interior of bubbles, the background spacetime continues to inflate with an expansion rate $H_F$ given by the false vacuum energy. As this space grows, new regions of inflating Hubble volumes become available and new bubbles can potentially nucleate
on these regions. The growth and bubble production process continues to future infinity. Although the number of bubbles produced is infinite, the volume occupied by them is only a fraction of the inflating background vacuum.
The general understanding has been that the phase transition never completes and such an evolution should lead to a nearly 'steady-state' of future-eternal inflation.

The solution to the Einstein Equations, $G_{a b} = \kappa T_{a b}$, for a spacetime with vacuum energy $\Lambda = \kappa^{-1} H_{F}^{2}$ is a global DeSitter (DS) geometry. The inflating background spacetime covers only half of this DeSitter (DS) space. The other half of the DeSitter space is a contracting spacetime, where the universe starts large and then contracts to size zero.  In eternal
inflation scenarios the contracting patch must be cut off in order for inflation to be successful. The separation of phases is achieved by imposing a hard boundary, the surface of the initial conditions (I.C.) passing through the throat of DS.

The global DS spacetime is covered by the metric

\be
ds^2  =  d{\tilde t}^2 - \frac{1}{H^{2}_F} \cosh(H_F {\tilde t})^2 d\Omega^{2}_{3} 
\label{fullds}
\ee

where $d\Omega_3$ is the standard metric of a $3-$sphere with coordinates ($r, \theta, \phi$). The contracting phase is covered by ${\tilde t} < 0$, followed by the expansion phase covered by ${\tilde t} > 0$. Bubble collisions and consequent thermalization of the
whole spacetime, leading to a completed phase transition, in the contracting phase have probability 1. For this reason, the contracting patch of DS space is cut off from the inflating phase by placing the initial conditions (I.C.) at
the boundary ${\tilde t} = 0$ where the two phases meet.

With such a setup, inflation begins at one particular special point at time ${\tilde t} = 0$ but it is believed that it never ends, i.e. it continues to future infinity. A useful metric that covers only the expanding phase
${\tilde t} > 0$ of DS space, (the inflationary phase), can be obtained through a coordinate transformation, given by flat coordinates

\be
ds^2  =  dt^2 - e^{2 H_F t} \left( dr^2 + r^2 d\Omega_{2}^{2} \right)
\label{flatds}
\ee

where $d\Omega_2$ is the metric of a $2-$sphere with coordinates ($\theta, \phi$). Notice that the initial conditions boundary ${\tilde t} = 0$ is pushed to $ t = -\infty$ in the new coordinates through the coordinate transformation. The scale factor $a(t) = e^{H_{F} t}$ is zero at the I.C. surface $a_{i} = a(t_{i} =-\infty) = 0$.

We can take $H_F = 1$ in Planck units without loss of generality.
The flat DS space can be embedded in a $5 D$ Minkowski flat space given by \cite{GGV}

\be
{\vec X}^2 + W^2 - V^2 = 1 
\label{mink}
\ee
where: $V = \sinh(t) + \frac{e^{t} r^2}{2}$,  $W = \cosh(t) -\frac{e^t r^2}{2}$ and ${\vec X} = e^t {\vec x}$. The I.C. surface is found by the condition $ V + W = a_{i} = 0 $. (For the global DS metric, the embedding is given by the transformation: $V = \sinh ({\tilde t}), W = \cosh( {\tilde t}), X_{i} = (X, Y, Z) = \cosh({\tilde t}) x_{i}, x_{i} = (x, y, z)$ ).

Bubbles are assumed to nucleate at negligible size and wall thickness. Since they grow with the speed of light their initial size is irrelevant to the discussion here. The interior of a bubble, nucleating for example at $t = r = 0$, is covered by the   
metric of the open FRW line element
\be
ds^{2}  =  d{\tau} ^2 - \sinh({\tau})^2 \left( d\xi^2 + \sinh({\xi})^2 d \Omega^{2}_{2} \right)
\label{interiorbubble}
\ee
where $\tau$ and $\xi$ are obtained from the original coordinates via the
transformation $V = \sinh (\tau) \cosh(\xi)$,  $W  = \cosh(\tau)$,  $X_i =
\sinh(\tau) \sinh(\xi) n_i$ with $n_{i}$ the coordinates that describe the standard metric of $d\Omega_{3}$. The Penrose diagram of the DS spacetime with the bubble nucleation and the observer's worldine are shown in Fig.1.

If an observer is positioned at some $\xi = \xi_{obs}$ along $Z$ we can always   
bring this point to $\xi'_{obs} = 0$ by using the Lorentz boost to transform to the observer's frame. The new coordinates in the boosted frame are
\bea
V' = \gamma \left(V  - \beta Z \right),\  Z' = \gamma \left(Z - \beta V \right) \nonumber\\
 {\rm and,} \,\ X'_{i}  = X_{i},     W'  =  W \nonumber
\eea
where the relativistic boost factor $\gamma  = \frac{1}{ ( 1 - \beta^2 )^{1/2}} $ is $\gamma = \cosh(\xi_{obs})$ and velocity $\beta = \tanh(\xi_{obs})$.

The key point for our purposes, elaborated in Sec.3, is that in its boosted frame the observer at  
$\xi'_{obs} = 0$ will see the initial conditions surface $a_{i} = a(t_i ) =
0$ tilted to a new position $a'_{i} = a'(t_i) < 0$ such that this surface
now cuts through and occupies portions of the contracting phase \cite{GGV}.

With $\lambda$ the nucleation rate per unit time and unit volume, let us denote  by
$V_{4}( \xi_{obs}, \tau, a_{i} )$ the 4-volume of spacetime sandwiched between the volume of the past light cone
of the observation point (point $P$ in Fig.1) and the past light
cone volume of the nucleation point of our bubble (point $N$ in Fig.1), modulo the interior section of the bubble $PP'N$ between these two light cones. This is
the volume available for nucleations of bubbles that can collide with ours, ( represented by the volume $ NP'P''N'$ in the Penrose diagram, Fig.1)  experienced by the observer at $P$.
The probability that no bubbles have collided with the observer in our bubble (located at $ \xi_{obs} $ ) before some proper time $\tau$ is
\be
{\cal P}  =  e^{ - \lambda  V_{4} }
\label{probab}   
\ee
The probability of the observer having 'seen' collisions is thus ${\tilde{\cal P}} = (1 - {\cal P}) $. Then the probability per unit time and per unit solid angle that the
observer at $P$ will 'see' a collision before time $\tau$ is
\be
\frac{d \tilde{{\cal P}}}{d\tau d\Omega'} \simeq \frac{\lambda d V_{4}}{d\tau d\Omega'}
\label{diffprob}
\ee
A number of authors \cite{GGV, Bucher} found that this
distribution of collisions is anisotropic
\be
\frac{dV_{4}}{d\tau} \simeq \gamma F\left(\xi_{obs}, \tau \right) \nonumber
\label{anisot}
\ee
where $F( \xi_{obs}, \tau)$, a function of $\xi_{obs}, \tau$, reduces to $F\simeq \frac{4\pi}{3}$ for $\tau \simeq 0$. 
The anisotropy per unit solid angle depends on the position of the observer $(\theta',
\xi_{obs})$ and generally on the observer's proper time $\tau$ from the initial surface. Notice that the $4-$volume diverges when the boost $\gamma$ becomes large.

The difficulties that the anisotropy may cause in achieving the homogeneity and
isotropy we observe in our universe have been known for a while \cite{GGV, Bucher, BordeG}. Below we explore a new and unexpected 
consequence of the anisotropy of bubble universes towards the surface of
the initial condition: reasons to question the very existence of eternal
inflation.

\begin{figure}[t]
\raggedleft
\centerline{\epsfxsize=2.8in
\epsfbox{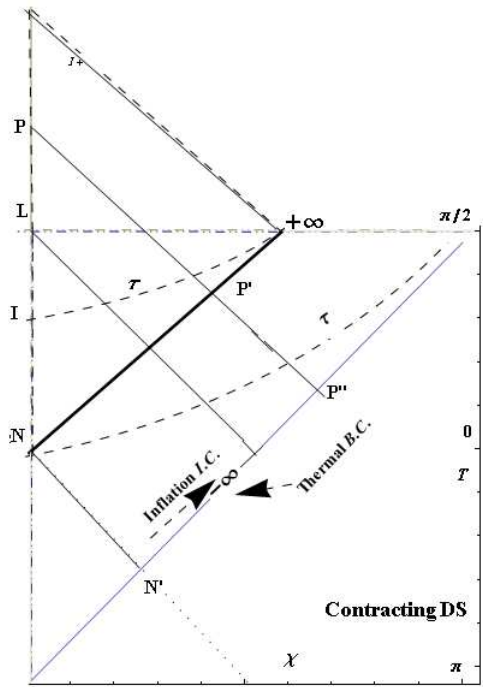}}
\label{fig:1}
\caption{Penrose diagram for a bubble nucleating at point $N$ with $t = r = 0$ in DS space, obtained from $\tan(T) =\cosh(\tau), \cos(\chi) = \cosh(\xi)$. The line $NP$ indicates the worldline of an observer, presently at $P$ in the bubble. Different cosmological epochs at hypersurfaces of constant proper time $\tau$ are also indicated, e.g.point $I$ marks the end of the second stage of inflation in the bubble's interior and $L$ the last scattering surface. Light cones are indicated by the pale thin lines, e.g. by line $PP'P''$ for an observer at $P$ inside the bubble. Hypersurfaces of constant proper time $\tau$ are indicated by the dashed curves, marked $\tau$. The observer's external $4-$volume $V_4$ available for nucleation of colliding bubble is $NP'P''N'$. The initial conditions surface is the diagonal line $W = -V$ i.e $t = - \infty$ with inflationary phase starting just above it and the end of the contracting phase just below it. The boundary condition of the contracting phase is that spacetime is all thermalized. At the line $W = -V$, this boundary meets the initial condition of the inflationary phase that impose that spacetime is all inflationary, thus the problem with Bianchi identity.}
\end{figure}

\subsection{3. Past and Future Incompleteness of Inflation due to the Initial
Conditions}

An observer stationed at $\xi_{obs}$ or a bubble nucleating at that point,
will experience uniform acceleration relative to the preferred frame of the
background. The observer's velocity relative to the preferred frame is
given by $\beta = v/c$ with a Lorentz factor $\gamma =
\frac{1}{ (1-\beta^{2})^{\frac{1}{2} }}$. The observer's proper time, $\tau$, measured from the initial conditions surface, 
 is estimated from the Lorentz factor $\gamma$ \cite{BordeG} as
\be
\tau  =  \frac{1}{2} ln \left(\frac{\gamma + 1}{\gamma - 1} \right)
\label{propertime}
\ee
Observers have their velocities relative to the comoving geodesic observer, 
redshifted by the scale factor $a(t)$ such that $v \simeq \frac{1}{a(t)}$. 
In the far future (large $t, \tau$ ) they align with the comoving observers of the geodesics for 
$t\rightarrow +\infty$ since $v\rightarrow 0$ . The Lorentz factor with respect to the preferred
frame can be found from Eqns.(\ref{flatds}, \ref{interiorbubble}) through the expression, $\gamma = \frac{dt}{d\tau}$, which for small $\tau
\rightarrow 0$ becomes $\gamma \simeq \cosh(\xi_{obs})$. From Eqn.(\ref{propertime}) it
can be seen that in the limit of a large proper time $\tau \rightarrow +\infty $ from the I.C. surface, the Lorentz factor tends to its minimum, $\gamma \rightarrow 1$, because 
in the limit $\tau \rightarrow + \infty $, the velocity of 
bubbles nucleating or observers located far from the initial conditions
surface relative to the preferred frame vanishes, $\beta = \frac{v}{c} \rightarrow 0$. From $\beta^2 = (1 - \frac{1}{\gamma^2} ) \simeq 0$ we thus have $\gamma \simeq 1$. 

The probability of collisions per unit time and unit angle \cite{GGV,Bucher} is proportional
to the differential spacetime $4-$volume, (Eqn.(\ref{diffprob}) )
\be
\frac{\lambda dV_{4}}{d\tau d\Omega'}  =  \frac{\lambda}{3} \gamma \left( 1 + \beta cos\theta')
\label{yll}  
\right)
\ee

and it is clearly anisotropic. Here $\theta'$ indicates the direction of observation in the boosted coordinates.   
The anisotropy towards the initial conditions surface and in the observed 
distribution of bubble nucleations and collisions \cite{GGV,Bucher}, depends on the location of the observer, $(\xi_{obs},
\theta', \tau)$ since the location of the observer determines the boost factor $\gamma$. The anisotropy disappears only in the limit of large proper time $\tau \rightarrow +\infty$, since the vanishing velocities $\beta \simeq \frac{1}{a(t)} \simeq 0$ there (discussed above), align the boosted frame with the background preferred frame.

We are interested to find out what happens to the anisotropy in the limit of small proper time, $\tau \simeq \epsilon \ll 1$.
At this stage, it important to notice that although the boost factor is bound from below in the limit of large proper times, ($\gamma \simeq 1$ and $\beta \simeq 0$ for $ \tau \simeq \infty$), the boost $\gamma$ becomes unbounded from above. With this in mind, let us investigate the regime where the boost, $\gamma$, may diverge: the regime of  
observers or bubbles nucleating near the initial conditions surface, $\tau \ll 1$. The initial conditions for eternal inflation at $t = t_{i} = -\infty$ where the scale factor $a_{i} = a(t_i ) = 0$,
make the assumption that no bubbles nucleate on the surface $a_{i} = 0$ or $\tau = 0$.
However bubbles can start nucleating at some very small proper time, just   
$\epsilon \ll 1$ away from the initial conditions surface, $\tau = 0 +
\epsilon \ll 1$. Such initial conditions are artificial and lead to inconsistencies of the theory as explained in Sec.4. The key point here is that from Eqn.(\ref{propertime}), observers stationed only a small proper time from the initial conditions surface, clearly have unbounded boost factors  
$\gamma \rightarrow +\infty$ and large velocities $\beta \rightarrow
1$, since in the limit $\tau \simeq \epsilon \ll 1$ the boost is given by

\be
\gamma \simeq  1 + \frac{1}{\epsilon} 
\label{gamma}
\ee

Thus $\gamma \rightarrow \infty$ for $\epsilon \rightarrow 0$.
 Physically, the large velocity  and boost are due to the blueshifting
effect from geodesics convergence, as the initial conditions singularity is approached (for $\tau \rightarrow 0$). For this reason, a bubble nucleating, for example along $Z$, at some small proper time
$\tau = 0 + \epsilon$ distance from the initial conditions surface, will have a   
a very large Lorentz boost $\gamma \simeq \cosh(\xi_{obs}) \rightarrow +\infty$
and large velocity $\beta \simeq \tanh(\xi_{obs}) \rightarrow 1$. {\it A diverging relativistic boost factor $\gamma$ leads to a diverging $4-$volume in Eqn.(\ref{yll}), therefore to a probability one of bubble destruction from Eqn.(\ref{diffprob}) and the end of eternal inflation} .

This problem arises from the fact that observers located near the initial conditions surface, with their large relativistic boost factor $\gamma$, experience a highly tilted initial conditions surface $a'_{i}$ in
their boosted frame \cite{GGV, Bucher}. In fact, their tilted I.C. surface $a'_i$ can be negative  $a'_i \le 0$. The nearer the observer is to the initial surface, then the larger their boost $\gamma$ and velocity $\beta$ are. But, as shown in Eqn.(\ref{negative}) below, the larger their boosts and velocities then the more negative values their tilted I.C. surface  $a'_i \le 0$ scans. Negative values of $a'$ simply mean that, in the observer's boosted frame, the I.C. surface $a'$ cuts below the original initial conditions $a_i = 0$ boundary of eternal inflation that separated inflating from contracting phases of the global DeSitter geometry. Observers with the tilted initial conditions surface, such that $a' < 0$, thus invade portions of the thermalized regions from the contracting DS spacetime, which were 'forbidden' by the inflationary cutoff $a_i =0$. Since observers near the initial conditions surface $a_{i} = 0$ have larger boosts $\gamma$ then they cover larger volumes of the thermalized spacetime region originally cut off from the inflationary chart, than the faraway observers with vanishing $\beta$'s and small boosts $\gamma \simeq 1$ . The relation between the boost factor of the observer $\gamma$ and the volume of the noninflationary DeSitter region being scanned by them, is problematic. This relation can be quantified by recalling that the initial conditions surface at ${\tilde t_i} = 0, t_i = - \infty$, that separates the inflationary phase from the contracting phase in the DS spacetime, is given by the constraint $ V + W = a_{i}$. In the observer's boosted frame with coordinates 

\be
V'  =  \gamma \left[ V - \beta  Z \right], \ Z'  =  \gamma \left[ Z - \beta  V \right]\nonumber
\label{tilta}
\ee

the initial conditions surface $a'(t_i)$ 'seen' by the observer becomes

\be
a' = V' + W'  =  -\beta Z' - \left(\gamma^{-1} - 1 \right) W' \le 0
\label{negative}
\ee

which is obtained from $ \gamma [V' + \beta Z' ] = -W'$. It can be seen from Eqn.(\ref{negative}) that in its boosted frame, the observer's past light cone occupies  
 a portion of the contracting DS spacetime below the boundary
 for eternal inflation $a_{i} = 0$, (Fig.2.a), which was originally cut off by the initial conditions 
boundary $W = - V$. So, the larger the observer's velocity $\beta$, the more tilted $a'_i$ becomes, implying that the more of the contracting spacetime is invaded by the observer's frame. But, larger velocities relative to the preferred frame correspond to observers and bubbles located near eternal inflation's surface of the initial conditions $\tau \simeq 0$ from $a_{i} =0$. In short, the closer an observer is to inflations's initial conditions, the more of the 'forbidden' spacetime region below the inflationary boundary their chart occupies.

The tilting of the initial conditions surface and the DS geometry as seen by the boosted frame are illustrated in Fig.2. The hyperboloid of Fig.2.a shows the global DS geometry obtained from Einstein equations. The contracting and expanding phases are separated at ${\tilde t_i} = 0 , t_i = -\infty $ by the initial conditions surface $a_{i} = e^{ t_i} = 0$ indicated by the diagonal plane in Fig.2.a. DS geometry as seen by the boosted observer is shown in Fig.2.b. It can be seen that the initial conditions plane $a'_{i} \le 0$ in the boosted frame is now tilted to a new position. In Fig.2.c the DS spacetime of case $(b)$ seen by the observer in the boosted frame is superimposed to the global DS geometry of Fig.2.a in order to compare how much of the thermalized region the boosted observer's chart invades. The two global phases in Fig.2.c are colored, red for the inflationary half of the spacetime and blue for the contracting part. For the case of a boost with $\beta \simeq 0.9$ depicted in Fig.2, it can be seen that the tilted I.C. plane $a'_{i}$ cuts below the boundary of inflation $a_i =0$ and covers a large part of the contracting spacetime (blue). The boosted observer can thus come in contact with  the 'forbidden' (blue) thermalized regions of spacetime, initially cut off from the inflationary chart via the I.C. boundary $a_i = 0$.

Why is the anisotropy towards the initial conditions, experienced by the observers as $a'_{i}\le 0$, problematic to the continuation of inflation? As we now demonstrate, due to $a'_{i} < 0$, inflation can not be future-eternal, instead it ends soon after the first bubbles that form near the initial conditions surface. The trouble comes from the fact that the volume of spacetime below the global I.C. surface $a_{i} = 0$, is completely thermalized with no inflationary regions left since it corresponds to the end of the contracting phase of DeSitter (DS) geometry. Towards the end of
the DS contracting phase, (just below the $a_{i} = 0 $ boundary), spacetime has
contracted to its minimum size near the boundary, all the bubbles
have merged, have grown to fill the whole spacetime, and thermalized. From Eqn.(\ref{propertime}) we know that bubbles forming near the I.C. surface, with $\tau \simeq 0$, have unbounded relativistic boosts $\gamma \rightarrow +\infty$ and large velocities $\beta = \frac{v}{c} \rightarrow 1$. But in this limit $\gamma \rightarrow +\infty$, their $4-$ volume per unit time and solid angle diverges when the boost is large $\gamma \simeq +\infty$ as follows from Eqn.(\ref{yll}). From Eqn.(\ref{probab}), a diverging volume means that their probability to get hit by the thermalized regions and other bubbles is one. A diverging spacetime $4-$ volume of the highly boosted observers $\gamma \gg 1$, implies that the boosted frame $a'_{i} <0$ {\it occupies too much of the contracting DS phase} in the global DS spacetime. Consequently, the first bubbles that form near the I.C. surface soon after the onset of infation, collide and are destroyed immediately upon formation, with probability one. 

All observers near the initial conditions region (with proper time $\tau \simeq 0$) have diverging 
boosts and velocities,  $\gamma \rightarrow +\infty, v\rightarrow c$, as can be seen from Eqn.(\ref{gamma}) for the limit $\tau \simeq \epsilon \ll 1$ in Eqn.(\ref{propertime}). Therefore they have highly tilted initial conditions surfaces $ a'_{i} < 0$ allowing them to invade too much of the thermalized region below the onset of inflation. According to Eqn.(\ref{negative}) then, all bubbles near the onset of inflation get hit with other bubbles upon formation and with the thermalized regions originally not covered by the eternal inflation spacetime, (regions below the initial boundary $a_{i} = 0$), resulting from Eqn.(\ref{yll}) and Eqn.(\ref{probab}). Then inflation ends soon after the onset, and eternal inflation becomes unlikely to be realized. As can be seen, the problems stemming from the choice of initial conditions in these scenarios are in close analogy with the transplanckian problem of Hawking radiation in which the frequency of the wavepackets is infinitely blueshifted near the horizon. 

Such an instability of the theory, the end of eternal inflation, is a direct consequence of the choice of the initial conditions, and it reflects the underlying nonlocal relationship between the preferred frame (seen as $a'_i$ by the observer) and the inflationary metric, Eqn.(\ref{negative}) with its initial conditions (fixed at $a_i = 0$).

\subsection{4. Discussion}
Let us probe into the origin of these unexpected difficulties in achieving future eternal inflation. Physically, introducing a cutoff in the theory by imposing the inflationary initial conditions at some special time-slice, $W = -V$ or equivalently $a_{i} = 0$, leads to a preferred frame that breaks Lorentz invariance. More importantly the stationary ($t=t_{i}$) preferred frame of the inflationary background breaks the general covariance of the theory \cite{ted}, i.e. the consistency of the Einstein Equations. As a result, observers with unbounded Lorentz boosts $\gamma$ positioned near the initial conditions hypersurface, can scan portions of the 'forbidden' contracting part of the DS spacetime below the I.C. boundary $W + V < 0$. That part of the global spacetime originally separated from the inflationary region by imposing the initial conditions boundary, is all thermalized. With probability one, bubbles near the initial conditions region, with small proper distances $\tau \simeq \epsilon \ll 1 $ thus large boosts $\gamma \simeq \frac{1}{\epsilon}$, Eqn.(\ref{gamma}), invade the thermalized regions originally exluded from the inflationary spacetime via the boundary $a_i$. Immediately upon formation they get destroyed and inflation ends. The root of the problem here lies in the breaking of general covariance of this theory by the I.C. of eternal inflation: near the I.C. surface, Einstein Equations $G_{a b} = \kappa T_{a b}$ are not satisfied since the the divergence of the Einstein tensor and the stress-energy tensor do not vanish, i.e. Bianchi identity is not satisfied. This can be seen by integrating the covariant energy conservation equation  
\be
{\dot \rho_{DS}} + 3 H \left(\rho_{DS} + p_{DS} \right) = 0 
\ee
around the boundary $a_i = 0$ of inflation's initial conditions for $\tau \simeq \pm \epsilon$. Here $\rho_{DS}(\pm \epsilon)$ denotes the vacuum energy of the respective DeSitter regions, a small distance above and below the initial conditions $a_i$ at ${\tilde t} = 0$. 
But due to the boundary $W = - V$ at ${\tilde t} = 0$ we have
\be
 \frac{\delta \rho_{DS}}{\delta t} \simeq \frac{\rho_{DS} (+\epsilon) - \rho_{DS} (-\epsilon)}{2 \epsilon} \ne 0
\label{energy}
\ee
The energy difference in Eqn.(\ref{energy}) goes to a finite value since there is a jump discontinuity in the energy values of the two phases, the inflating phase with energy $\rho (+\epsilon) \simeq H_{F}^{2}$ just above and at the boundary $a_{i} =0 $ and, the energy of the thermalized phase $ \rho(-\epsilon) \simeq 0 $ just below the boundary $a_i$ of the I.C. surface. So, Bianchi identity is not satisfied in the limit $\epsilon \rightarrow 0$ near the initial conditions region. The inconsistency of Einstein equations in eternal inflation is a direct consequence of the choice  of the initial conditions at ${\tilde t} \simeq 0$ or $t \simeq -\infty$ that make the assumption that there are no bubbles at the onset of inflation, i.e. spacetime is completely inflationary there, although only an infinitesimal proper distance $ (-\epsilon)$  just below the initial conditions time-slice, spacetime is all thermalized. Inevitably these initial conditions lead to difficulties in actualizing eternal inflation. So far we have treated the problem at the classical level. An interesting question is whether fluctuations around the initial conditions may remedy the breaking of general covariance \footnote{We thank the referee for raising this point}.  Although the issue of fluctuations for the bubbles and the background will be rigorously treated in a separate paper, theorems proven in \cite{BordeG, BordeV, Borde, guth1} and the analysis of \cite{ted} are sufficient to indicate that fluctuations around the initial conditions singularity are likely going to be out of control for the following reasons: we are dealing with an initial singularity, the point where the congruence of geodesics converge. Thus if spacetime can not be extended beyond this point, it is not clear how one can extend the fluctuations; these scenarios generically break energy conditions, which imply unstable fluctuations; they contain a stationary preferred frame which leads to incurable pathologies and nonlocal fluctuations \cite{ted}. 

It would be interesting to explore whether  a different set of initial conditions, such as assigning a surface gravity or tension to the boundary $a_{i} = 0$ that satisfies Bianchi identity, may produce viable models of eternal inflation. Until recently it was thought that inflation's initial conditions were decoupled from our bubble universe, i.e. that eternal inflation leads to a case where the only physics relevant for the cosmology of a bubble universe is that of the inflating region far above the I.C. surface. That view is of course based on the existence of inflation to future-eternity. If inflation were eternal then it makes sense to imagine our bubble universe being formed safely, far away from the initial conditions surface (i.e at $a(t)\gg 0, \beta \simeq 0$) and the problematic thermalized regions below that surface, such that the only lasting effect would be a vague memory of the initial surface \cite{GGV}. Contrary to that belief, we have shown here that the initial conditions problem remains relevant to eternal inflation. Eternal inflation seems very difficult to achieve with these initial conditions. They lead to a background preferred frame which breaks the general covariance of the theory, thereby leading to instabilities of a 'steady state' inflation, and to the inconsistency of Einstein equations. Imposing an 'artificial' boundary on the metric at the onset of the inflationary phase, can not prevent bubbles nucleating after the onset of inflation to collide and access the contracting spacetime with probability one. Could it be inevitable that the beginning of eternal inflation marks an ephemeral end?



\begin{figure}[t]
\raggedleft
\centerline{\epsfxsize=3.2in
\epsfbox{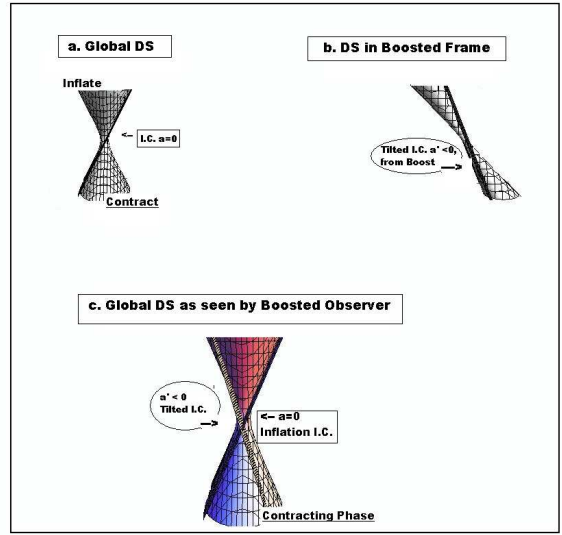}}
\label{fig:2}
\caption{(a) Global DS Spacetime with the Initial Conditions surface $a_{i}$ separating the Inflating phase (upper half) from the Contracting phase (bottom half). (b) The second geometry indicates the global DS geometry in the boosted frame of the observer. Notice the tilted initial conditions surface $a'_{i}$ the observer 'sees'. (c) The third DS spacetime depicts the observers tilted boundary $a'_{i}$ relative to $a_{i}$. Inflationary phase in (c) is in red and contracting phase in blue and the $45^0$ plane is $a_i = 0$ separating the two phases. We can see in (c) how much of the contracting (blue) DS phase the observer's boosted frame covers due to the initial boundary $a'_{i} \le 0$ piercing below the inflation's initial conditions $a_{i} = 0$. The diagram is plotted for the representative value $\beta \simeq 0.9$. }
\end{figure}



\begin{acknowledgments}
LMH thanks CITA for their hospitality and D.Bond for helpful discussions. LMH was suported in part by DOE grant DE-FG02-06ER1418, and the Bahnson fund.
\end{acknowledgments}

\subsection{Appendix}

The approach sometimes taken in literature, in relating the landscape of string theory to eternal inflation, can be confusing. The ensemble of bubble universes that would be produced by eternal inflation does not require a landscape to exist, since it is supposed to create its own through the bubble production. On the other hand, the landscape of string theory can and does also exist independently of the universes inflationary cosmology. There are many ways to populate the string landscape, for example by the wavefunctions of the universe proposal \cite{us1}. Of course it would be formidable if these two theories could somehow be unified. Current efforts along these lines exist in literature. However, a common misconception in most of these efforts is to study inflation on the string laqndscape as a double-well problem, i.e. to apply the formalism of a two-body system to an $N-$body problem. Tunneling between false and true vacuum and interpolation of the field between these two vacua is the common theme in many eternal inflation models. On the other hand, the landscape of string theory contains a large number, infinite for all practical purposes, of vacua, with their own intricate structure derived from string theory. Commonly, efforts to place the inflaton field and incorporate eternal inflation on the landscape of string theory, continue to approach and calculate the evolution of the inflaton field as a tunneling process between two neigbhoring vacuum sites. This is completely incorrect. The physics of a scalar field interacting with an $N-$body lattice of vacua, such as the landscape, is totally different and can not be reduced to the simple two-body problem of a barrier separating false vacuum from true vacuum. Let us think of the analogy with a condensed matter system: the situation where an electron interacts with only two atoms is very different from an electron interacting with a wire with an almost infinite number of atoms. If we were to oversimplify and reduce the case of the electron on the wire to the case of the electron interacting with only two atoms, then we are guaranteed we would get the wrong answer for the electron's field, probability current and its emerging behaviour, such as conductivity or localization. By the same token, reducing the N-body problem of an inflaton field on the landscape to the problem of an inflaton field interacting with only two neighboring potential wells, is an oversimplication which leads to an incorrect answer. The correct physics for an $N$-body system can not be captured by a 2-body system, as is well known. A beautiful description of the difference between the two-body and the $N-$body physics, along with the perturbation theory methods for the $N-$body systems, is given for example in \cite{anderson}. Besides the issues this problem has in common with a condensed matter $N-$body system, in cosmology addressing decoherence in an interacting $N-$body system becomes essential. It would be interesting to see how eternal inflation is accomodated on the landscape framework, when studied thoroughly by including a decoherence mechanism and by applying the $N-$body physics formalism.

\end{document}